\documentclass{webofc}

\usepackage[varg]{txfonts}   % Web of Conferences font
\usepackage{graphicx}% Include figure files
\usepackage{dcolumn}% Align table columns on decimal point
\usepackage{bm}% bold math
\usepackage[mathlines]{lineno}% Enable numbering of text and display math
\usepackage[utf8]{inputenc}
\usepackage[bottom]{footmisc}
\usepackage[T1]{fontenc}
\usepackage{mathptmx}
\usepackage{subcaption}
\usepackage[colorlinks,urlcolor=blue,citecolor=blue]{hyperref}
\fancyhfoffset[L]{1cm} % left extra length
\fancyhfoffset[R]{1cm} % right extra length
\begin{document}

\vspace{50pt}

\begin{center}
\vspace{20pt}
    \fontsize{15pt}{15pt}\selectfont
        %\textsc{\textbf{Neutron Star Mergers and the Quark Matter Equation of State}}
        %\vspace{50pt}
         \textbf{  ~ \\
         ~ \\
         ~ \\
         Neutron Star Mergers and the Quark Matter Equation of State}
\end{center}

%\title{Neutron Star Mergers and the Quark Matter Equation of State}

\vspace{12pt}

%\author{\firstname{Grant J.} \lastname{Mathews}\inst{1}\fnsep\thanks{\email{gmathews@nd.edu}} \and
%        \firstname{Atul} \lastname{Kedia}\inst{1}\fnsep\thanks{\email{atulkedia93@gmail.com}} \and
%        \firstname{Hee Il} \lastname{Kim}\inst{2}\fnsep\thanks{\email{khizetta@sogang.ac.kr}}  \and
%        \firstname{In-Saeng} \lastname{Suh}\inst{1,3,4}\fnsep\thanks{\email{isuh@nd.edu; suhi@ornl.gov}}
        % etc.
%}

%\institute{Department of Physics and Astronomy, Center for Astrophysics, University of Notre Dame, Notre Dame, IN 46556, USA
%\and
%           Center for Quantum Spacetime, Sogang University, Seoul 04107, Korea
%\and
%           Center for Research Computing, University of Notre Dame, Notre Dame, IN 46556, USA 
%\and
%           National Center for Computational Sciences, Oak Ridge National Laboratory, Oak Ridge, TN 37830, USA
%          }
\noindent
{\it Grant J.} Mathews\footnote{gmathews@nd.edu}\inst{\ast}, {\it Atul} Kedia\footnote{akedia@nd.edu}\inst{\ast}, {\it Hee Il} Kim\footnote{khizetta@sogang.ac.kr}\inst{\dagger}, and {\it In-Saeng} Suh\footnote{isuh@nd.edu;suhi@ornl.gov}\inst{\ast}\inst{\ddagger}\inst{\mathsection}

%        \firstname{Atul} \lastname{Kedia}\inst{1}\fnsep\thanks{\email{atulkedia93@gmail.com}} \and
%        \firstname{Hee Il} \lastname{Kim}\inst{2}\fnsep\thanks{\email{khizetta@sogang.ac.kr}}  \and
%        \firstname{In-Saeng} \lastname{Suh}\inst{1,3,4}\fnsep\thanks{\email{isuh@nd.edu; suhi@ornl.gov}}
        % etc.
%}
\vspace{12pt}

\noindent
$^{\ast}${Department of Physics and Astronomy, Center for Astrophysics, University of Notre Dame, 
Notre Dame, IN 46556, USA} \\
%\and
$^{\dagger}${Center for Quantum Spacetime, Sogang University, Seoul 04107, Korea} \\
%\and
$^{\ddagger}${Center for Research Computing, University of Notre Dame, Notre Dame, IN 46556, USA} \\
%\and
$^{\mathsection}${National Center for Computational Sciences, Oak Ridge National Laboratory, Oak Ridge, 
TN 37830, USA} \\

{\bf  Abstract}  As neutron stars merge they can approach very high nuclear density. Here, we summarized recent results for
the evolution and gravitational wave emission from binary-neutron star mergers using a a variety of nuclear
equations of state with and without a crossover transition to quark matter.    We discuss how  the late time gravitational wave emission
from binary neutron star mergers may possibly reveal the existence of a crossover transition to quark
matter.

%\maketitle
%
\section{Introduction}
\label{intro}
In recent work \cite{Kedia22} we have explored the effects of a crossover transition to quark matter on the emergent gravitational waves from binary neutron star mergers.  In this paper we summarize that work and other efforts toward unraveling the effects of the formation of  quark-matter during neutron-star mergers.   Neutron stars (NSs)  and NS binaries can probe the equation of state (EOS) at supra-nuclear densities (for recent reviews see Refs.~\cite{Baiotti19, Radice20}). Indeed, the detection of gravitational waves (GWs) from the GW170817 event by the LIGO-Virgo Collaboration \cite{LIGO-GW170817,LIGO-GW170817eos}  provided new insights into the properties of neutron-star matter \cite{Lattimer12}. Beyond that, determinations  of NS masses and radii by the NICER mission also constrain the EOS of nuclear matter \cite{Miller19, Riley21, Miller21}. Tidal effects can be inferred from the signal in  ground-based GW observatories \cite{Flanagan08,Hinderer08, Read09tidal}. In the LIGO-Virgo events tidal deformability ($\Lambda$) of a NS of mass $M=1.4~\rm{M}_\odot$ have also been inferred  $\Lambda_{1.4} < 800$ at (90\% C.L.) for a low-spin prior \cite{LIGO-GW170817} and the radius constraint for a $M=1.4 ~\rm{M}_\odot$ NS was deduced  to be $R_{1.4} < 13.6 ~\rm{km}$. Subsequently, this has been further constrained to be $R_{1.4} = 11.9 \pm 1.4 ~\rm{km}$ \cite{LIGO-GW170817eos}.   Also, newer constraints on the maximum NS mass and a  lower limit of the tidal deformability  were also inferred \cite{Most18,Annala18}. Adding the requirement that the equation of state  asymptotically approach the regime of  perturbative QCD \cite{Annala18, Kurkela10, Komoltsev22, Fujimoto22,Gorda22,Tootle22}, leads to constraints on the radius of a maximum-mass NS of $R_{\max} < 13.6 ~\rm{km}$ and $\Lambda_{1.4} > 120$ \cite{Annala18}. It has also been shown that an EOS with a phase transition can imply  $8.53 ~\rm{km} < R_{1.4} < 13.74 ~\rm{km}$  and $\Lambda_{1.4} > 35.5$ at the $3~\sigma$ level \cite{Most18}.

%The detected gravitational wave signal depends upon the tidal deformability of the NSs as they approach merger. 
%The tidal deformability \cite{Flanagan} was deduced following the LIGO analyses \cite{LIGO-GW170817,LIGO-GW170817eos}.
%implying that the radius of the stars of $1.4 ~\rm{M}_\odot$ is in the range $10.5 ~\rm{km} \le R \le 13.3 ~\rm{km}$. The inferred effective tidal deformability ($\Lambda$) of a neutron star with a mass of $M=1.4 ~\rm{M}_\odot$ was deduced to be $\Lambda_{1.4} < 800$ at the $90 ~\%$ confidence level \cite{LIGO-GW170817}. This constrains the radius to be $R_{1.4} < 13.6 ~\rm{km}$. Subsequently, this was further constrained to be $R_{1.4} = 11.9 \pm 1.4 ~\rm{km}$ \cite{LIGO-GW170817eos}. Subsequent analyses \cite{Annala18,Margalit17,Rezzolla18,Radice17,Ruiz18,Shibata17,Most18} has lead to further constraints on the maximum neutron star mass and tidal deformabilities. Of interest to the present work is that incorporating perturbative QCD at a density of $\simeq 40 ~n_0$ \cite{Kurkela10}, where $n_0$ is the nucleon saturation density, leads to $R_{\max} < 13.6 ~\rm{km}$ and $\Lambda_{1.4} > 120$ \cite{Annala18}. It has been also shown that equations of state (EOSs) with a first-order phase transition can give $8.53 ~\rm{km} < R_{1.4} < 13.74 ~\rm{km}$ at the 2$\sigma$ level and $\Lambda_{1.4} > 35.5$ at the $3\sigma$ level \cite{Most18}.

There is currently much interest in the fact that a phase transition in the EOS  can produce  a variety of dynamical collapse patterns (cf.~\cite{Weih20}). As explained below, such changes in the EOS can produce a shift of the maximum peak frequency ($f_{peak}$ (sometimes denoted as  $f_2$) in the detected power spectral density (PSD) \cite{Bauswein19, Blacker20, Radice17}. Such a  shift can violate the universal relation between $f_{peak}$ and  tidal deformability that has been noted for  pure hadronic EOSs \cite{Breschi19}. However, an EOS with a phase transition may not conform to the same empirical universal relations \cite{Bauswein12a,Hotokezaka13,Bernuzzi14,Rezzolla16,Zappa18}. Hence, an observed shift might indicate the formation of quark matter. This conclusion, however,  is  model dependent  (e.g. \cite{Most19, Most20}) and also depends upon the duration of  merger remnant  \cite{Weih20,Liebling21,Prakash21}.

 A number of recent works have discussed  EOS effects on the GW signal.  Some of them have also considered the formation of quark matter  \cite{Bauswein12b, Most20,Bauswein19,Bauswein20,Blacker20,Weih20,Liebling21,Prakash21}. Most of these studies, however,  have considered  a first-order phase transition.  In this case  a mixed quark-hadron phase forms which can  remove pressure support leading to a prompt collapse.  However, since the strength of the order parameter for the QCD phase transition is not known,  a simple crossover or  a weakly first-order  transition is possible \cite{Pisarski16, Steinheimer11, Hatsuda06, Aoki06, Baym18}. The  pressure in the regime of the crossover  could be large compared to a hadronic or  a first-order  transition. This could extend the  postmerger phase.   Hence, an observation of a long-duration  post-merger GW event, could  possibly indicate both the order of the transition and  the coupling strength of  quark-matter in the crossover regime \cite{Kedia22}.

In Ref.~\cite{Kedia22} we examined the crossover to the formation of quark-gluon plasma during the postmerger and demonstrated  that the GW signal from the postmerger phase is indeed sensitive to the quark-matter EOS. 
It was shown that that the properties of quark matter in the non-perturbative crossover regime of QCD increases  the pressure of the postmerger remnant. This leads to a longer  duration of the late time gravitational radiation such that the GW emission might become a means to probe the  non-perturbative regime of quark matter.

In particular, in Ref.~\cite{Kedia22}  various parameterizations of the quark-hadron crossover (QHC19) EOS  of \cite{Baym19} were  investigated. A  complementary  study has also been made in Ref.~\cite{Kojo21} based upon the newer (QHC21) version with similar conclusions.
As the density increases, a critical point is thought to appear.   Above that density a weak first-order chiral transition may occur \cite{Kronfeld12}.  In the QHC19 EOS the transition from hadronic to quark matter is treated as a continuous crossover  parameterized with a 5th order polynomial. The observational constraints on the  NS mass ($> 2 ~\rm{M}_\odot$) \cite{Demorest10,Antoniadis13,Fonseca21} and the  radius bounds from the LIGO-Virgo  analysis are satisfied in all versions of this EOS.

Within this context the tidal deformability, maximum chirp frequency $f_{max}$, and power spectral density  frequency peak $f_{peak}$ were analyzed in \cite{Kedia22} as a means to   identify observational signatures of the crossover to quark matter during binary NS mergers. The crucial postmerger GW emission occurs in a high  frequency range (1--4 kHz).  Although this frequency is outside the current LIGO/aVirgo/KAGRA window, it is anticipated that next generation of GW observatories such as the Einstein Telescope \cite{Sathyaprakash12} and the Cosmic Explorer \cite{Abbot17} will be sensitive in this frequency range. We argue that  observations of such higher frequency gravitational wave emission in the next generation detectors may have the possibility to characterize both the order of the transition and the physics of the  crossover regime of  quark matter.

\section{Equations of state}
\label{section:EOS}

At high baryon density and chemical potential  the QCD strong coupling  $\alpha_s$ approaches unity. A non-perturbative approach to QCD is then necessary to describe the generation of constituent quark masses,  chiral symmetry breaking \cite{Hatsuda94},  quark pairing, and color superconductivity \cite{Alford08}, etc.
For our studies we utilized various parameterizations of  the QHC19 EOS \cite{Baym19}. In that work, the low-density hadronic regime (i.e.  less that twice the nuclear saturation density,  $< 2 ~n_0$) utilized the Togashi EOS \cite{Togashi13,Togashi17}. This is an extended version of the relatively soft  APR EoS \cite{Akmal98}.  Our study \cite{Kedia22} instead utilized the SLy \cite{SLy} and the GNH3 \cite{GNH3} EoSs as bracketing the physics  of a soft and stiff EoS, respectively.

The QHC19 EOS accounts for the non-perturbative QCD effects n the context of the Nambu-Jona-Lasinio model (see Refs. \cite{Nambu61a,Nambu61b,Buballa05}). The Lagrangian contains four  coupling constants.   These are: 1)  the scalar coupling ($G$); 2) the coefficient of the Kobayashi-Maskawa-'t Hooft vertex ($K$); 3) the vector coupling for universal quark repulsion ($g_v$);  and 4) the diquark strength ($H$). In the QHC19 EoS, only  two coupling constants ($g_v$ and $H$, scaled to $G$) are varied  to construct versions of the model. The matter pressure increases as these couplings increase \cite{Baym19, Baym18}. In \cite{Kedia22} we utilized three parameter sets from \cite{Baym19}, identified as QHC19B [$(g_V,H) = (0.8,1.49)$], QHC19C [$(g_V,H) = (1.0,1.55)$], and QHC19D [$(g_V,H) = (1.2,1.61)$]. The pressure in the crossover regime ($2 ~n_0 < n < 5 ~n_0$) is given analytically with  fifth-order polynomials in terms of the baryonic chemical potential. The tidal deformability ($\Lambda < 800$ for $M_0 = 1.4 ~\rm{M}_\odot$) \cite{LIGO-GW170817}, the maximum mass \cite{Demorest10,Antoniadis13,Fonseca21}, and radius of neutron stars are all satisfied with these parameterizations of the QHC19 EOS.  For numerical speed we implemented the QHC19 EOSs using piecewise-polytropic  fits as described by Ref. \cite{Read09} and modified as discussed in  \cite{Kedia22}. 

\section{Simulations}
%\label{section:Code}

In \cite{Kedia22} binary merger simulations were evolved using the numerical relativity software platform, \texttt{the Einstein Toolkit} (ET) \cite{ET}. This platform incorporates  full general relativity in three spatial dimensions based upon  the BSSN-NOK formalism \cite{NOK,BSSN1,BSSN2,BSSN3,BSSN4}. The hydrodynamics was evolved with the use of the \texttt{GRHydro} code \cite{Baiotti,Hawke05,Mosta14} based on the Valencia formulation \cite{Valencia1,Valencia2}. The initial conditions were generated using \texttt{LORENE}\cite{LORENE,LORENE2}. The thorn \texttt{Carpet} \cite{Carpet1,Carpet2} was used for adaptive mesh refinement based upon  six mesh refinement levels and a minimum grid  of 0.3125 in Cactus units ($\approx 461 ~\rm{m}$). The thermal pressure component was implemented in \texttt{GRHydro} using a constant adiabatic index $\Gamma_{\rm{th}} = 1.8$  as in Ref.~\cite{Pietri16}.

The GWs emitted during the binary merger were extracted using the Newman-Penrose formalism which is based upon a multipole expansion of the spin-weighted spherical harmonics of the Weyl scalar $\Psi_4^{(l,m)} (\theta, \phi, t) = \ddot{h}_+^{(l,m)}(\theta, \phi, t) + i \ddot{h}_\times^{(l,m)}(\theta, \phi, t)$. The two polarizations of the  strain  $h_+(\theta, \phi, t)$ and $h_\times(\theta, \phi, t)$ were calculated by summing over the $(l,m)$ modes and integrated twice. 
%and then the amplitudes are scaled for binaries at a distance of $50$ Mpc from the observer.
The isolated NS models involved baryonic masses of $M_B =1.45$, $1.50$, $1.55 ~\rm{M}_\odot$, with similar gravitational masses $\sim 1.35-1.4$.  These were placed on the grid with an initial coordinate separation between centers of $45~\rm{km}$.

\begin{figure}
    \includegraphics[width=0.80\textwidth]{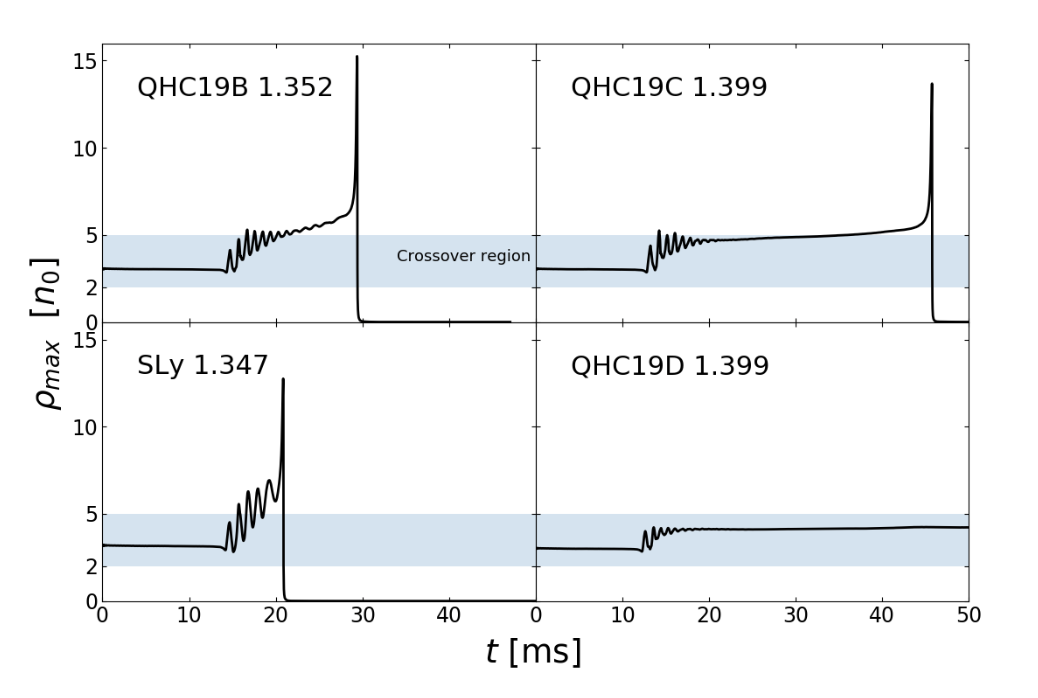}
    \caption{Evolution of maximum rest-mass density vs time for several equations of state from \cite{Kedia22}.  Numbers next to the EOS label indicate the gravitational mass of an isolated neutron star for each case. The blue band indicates QHC-crossover densities. In all cases, the NSs start in the crossover density range (2--5 $n_0$) followed by a rise in density, leading to a collapse to a black hole (in all except the bottom-right panel). The bottom-right case (QHC19D 1.399) does not form a black hole within the simulation time. }
    \label{fig:density}
\end{figure}

Figure \ref{fig:density} (from Ref.~\cite{Kedia22}) illustrates  the evolution of the maximum density during the simulations. This figure shows that the  densities in the NSs even before the  merger are well into the crossover range (2--5 $n_0$).  The NS core densities remain  in the crossover domain  at a densities of about $n \sim 2.95-3.15 ~n_0$ during the approach to merger. Subsequently, the maximum density rises until the maximum density exceeds $\sim 5-6 ~ n_0$. At this point  the core of the system collapses to the central black hole as evidenced  by a density spike in this figure.

%During the inspiral, the stars in the binary system tidally deform and start to coalesce. As such, $t_{inspiral}$ largely depends on the tidal deformability and the stiffness of the EOS at densities lower than the initial central densities. As the initial central densities of the QHCs lie in the range of $2.95-3.15 ~n_0$, this is where QHCs are close to but a bit stiffer than the SLy.

\begin{figure}
    \includegraphics[width=0.75\textwidth]{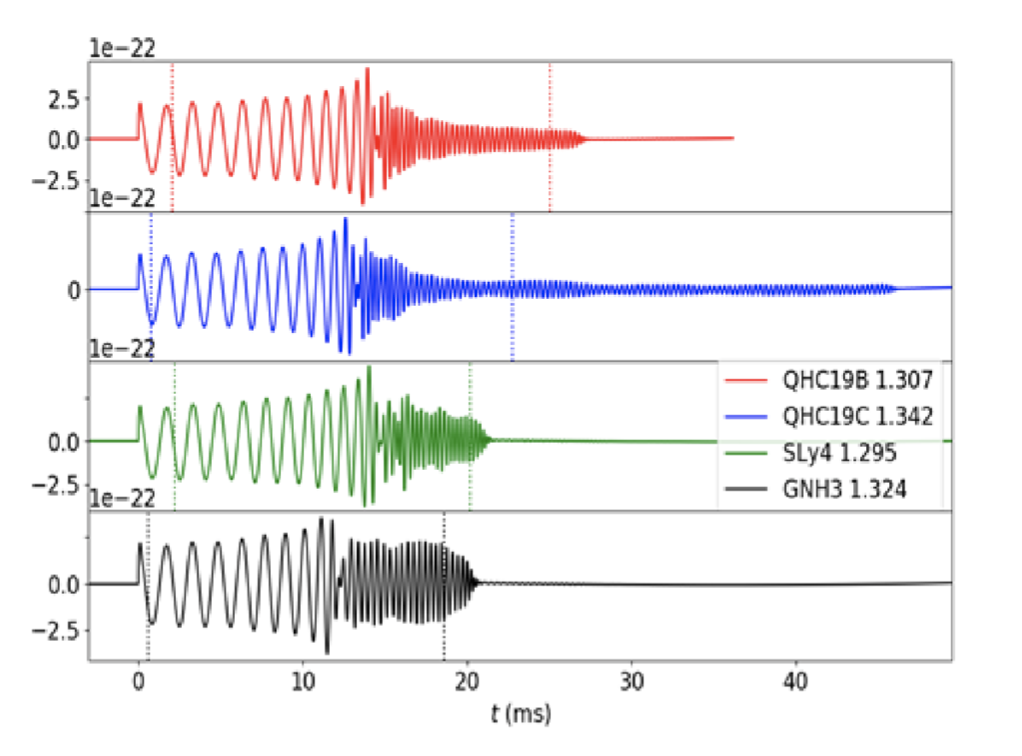}
    \caption{Evolution of the GW strain $h_{+,\times}$ vs time for several representative simulations with nearly identical starting conditions, but for different equations of state as labelled.  The numbers indicate the isolated neutron star mass for each EOS as indicative of the similarity of initial conditions.  The upper two curves are for parameterizations of the  QHC19 EOS, while the bottom two curves are for a soft and stiff pure hadronic EOS.  Note the the signal continues for a much longer duration in the cases with a crossover to quark matter. }
    \label{fig:GW}
\end{figure}

Figure \ref{fig:GW}  illustrates  the evolution of the strain for various equations of state as labelled, but for nearly identical initial conditions. The striking feature is that the GW signal endures for a much longer time for the cases with a QHC EOS.  Moreover, the larger the quark coupling, i.e. going from QHCB to QHCC, the longer the duration of the postmerger GW emission.   This suggests that one might learn the strength of the non-perturbative quark-matter couplings from the observation of an extended post merger phase.

Indeed, the postmerger duration, i.e., the lifetime of the hyper-massive neutron star (HMNS), strongly depends on the EOS stiffness at the crossover densities.   When  densities in excess of the nuclear saturation density are achieved in the core for the hadronic EOSs it is impossible to stop the merger remnant from collapsing into a black hole.  The postmerger remnants from binaries based upon the QHC19 EOS, however,  have sufficient pressure to delay gravitational collapse. As the EOS stiffness within the QHC models increases, longer lifetimes of their HMNS remnants are apparent. Even the  QHC19B EOS produces a  much longer postmerger duration than the  hadronic EOSs.  For the case of QHC19D, even the highest-mass case fails to  collapse.

\begin{figure}
    \includegraphics[width=0.75\textwidth]{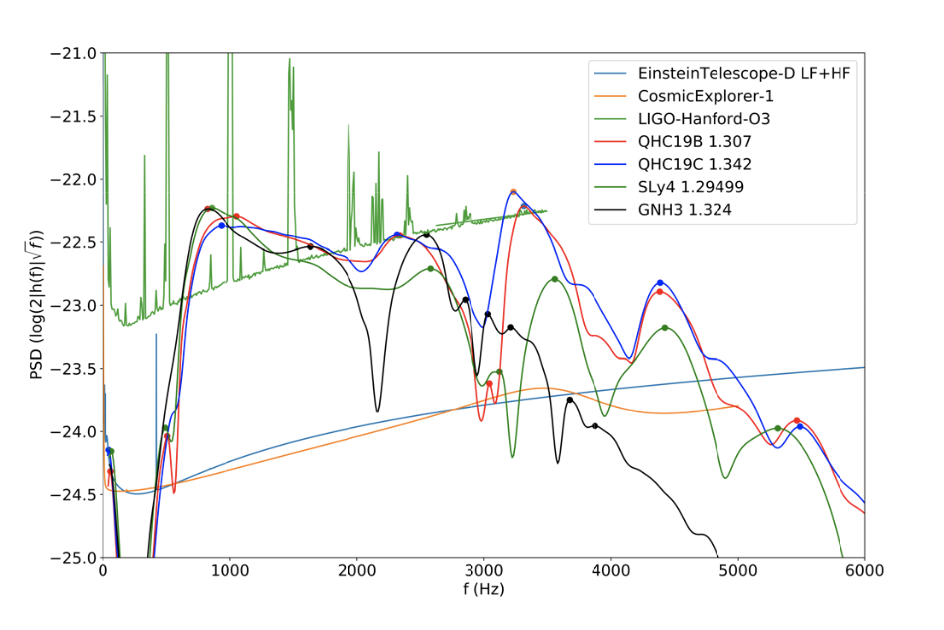}
    \caption{ Power spectral density ($2 \tilde h (f) f^{1/2}$) as a function of frequency for various simulations as labelled and shown on Figure \ref{fig:GW}. The upper green curve shows the LIGO sensitivity while the lower blue and orange curves show anticipated sensitivity of the Einstein Telescope and Cosmic Explorer, respectively.  The first peak at around 1 kHz for all of the simulations corresponds to the initial contact of the merging binaries.  The second peak near 2 kHz corresponds to the maximum chirp strain, $f_{max}$, while the third peak,  at around 3 kHz corresponds to the long postmerger phase, $f_{peak}$.}
    \label{fig:PSD}
\end{figure}

Of course, what is actually detected in GW observatories is not the strain, but its fourier transform.  In particular, an effective fourier amplitude can be deduced
\begin{equation}
\tilde h_{+,\times}(f) = \int h_{+,\times}(t) e^{-i 2 \pi f t} dt~~,
\end{equation}
This is usually plotted as a normalized  power spectral density (PSD) given by $2 \tilde h (f) f^{1/2}$ \cite{Takami15}.
Figure \ref{fig:PSD} shows some PSD spectra deduced from the simulations in Ref.~\cite{Kedia22}.   The upper green curve shows the LIGO sensitivity while the lower blue and orange curves show the expected sensitivity of the future Einstein Telescope and Cosmic Explorer, respectively.  The first peak at around 1 kHz for all of the simulations corresponds to the initial contact of the merging neutron stars, while the second peak near 2 kHz corresponds to the maximum chirp strain, $f_{max}=\frac{1}{2\pi}\frac{d\phi}{dt}|_{max}$, where $\phi$ is the phase of the strain (see \cite{Takami15}).   Of particular interest for probing quark matter, however, is  the third peak, $f_{peak}$,  at around 3 kHz corresponding to the long postmerger phase.  What can be noted in this figure is that the amplitude of $f_{peak}$ directly correlates with the duration of the postmerger system, and therefore, relates to the strength of the coupling constants in the QHC19 EOS parameterizations.

In spite of the promising feature that $f_{peak}$ PSD becomes large for a crossover to  quark matter, it is possible that other equations of state can lead to such a peak \cite{Takami15}.   What is needed, therefore,  is another unique signature to specifically identify quark matter.    In \cite{Kedia22} it was suggested that the softness of the QHC equations of state  at lower densities, $\sim 3 n_0$, is apparent  in their pre-merger $f_{max}$ frequency, whereas the stiffness at higher densities is indicated  in the postmerger $f_{peak}$ frequency. 

\begin{figure}
    \begin{subfigure}[b]{0.5\textwidth}
        \includegraphics[width=\textwidth]{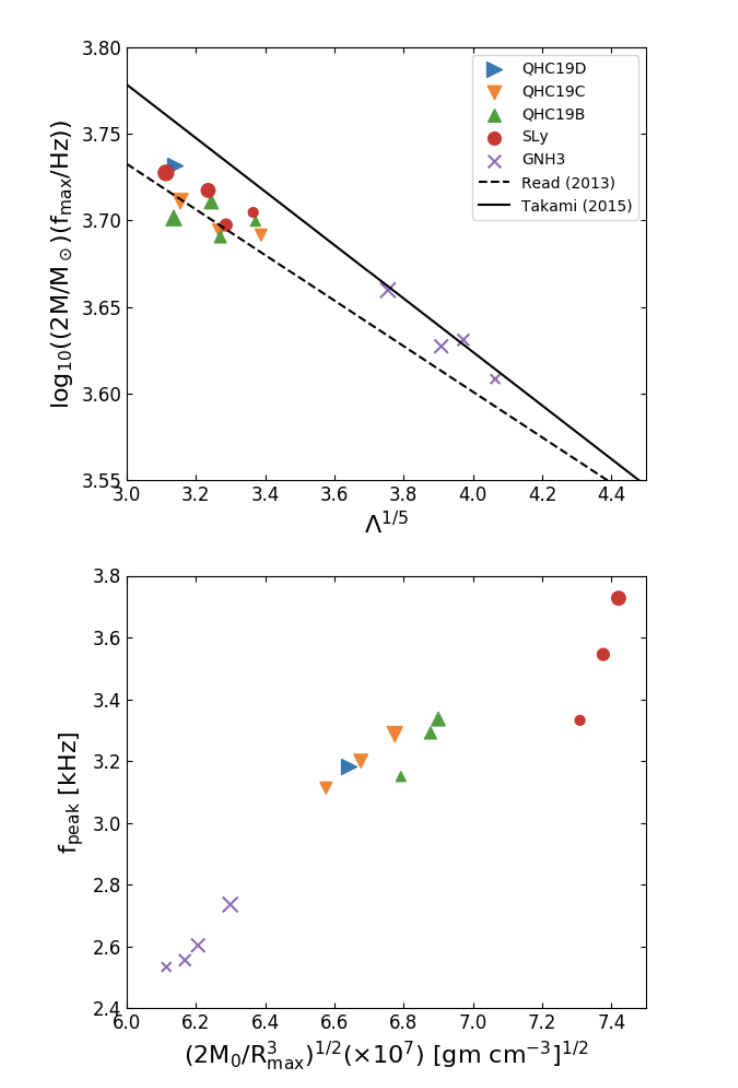}
   \end{subfigure}
%    \vspace*{-0.5cm}
    \caption{From Ref.~\cite{Kedia22}.  Top panel shows $f_{max}$ vs. the dimensionless tidal deformability ($\Lambda^{1/5}$).   Also plotted are universality relations suggested in previous work \cite{Read13, Takami15} as indicated. The lower panel shows $f_{peak}$ vs. the pseudo-average rest-mass density $(2M_0/R_{\max}^3)^{1/2}$. The size of the symbols indicates the size of  the gravitational mass in isolation $M_0$ as listed in \cite{Kedia22}.}
\label{fig:fmax}
\end{figure}

This dual nature of the QHC equations of state  (having both softness and stiffness) might  be revealed by observations of both $f_{max}$ and $f_{peak}$ in a single GW event.   This is illustrated in figure \ref{fig:fmax} from Ref.~\cite{Kedia22}.  The upper panel shows that $f_{max}$ values for the QHC equations of state in our simulations obey the scaling relations with tidal deformability as noted in \cite{Read13}.   This also shows that the QHC simulations all cluster with a soft EOS like the SLy in the initial chirp.  However, the lower panel shows the relation between $f_{peak}$ and the pseudo-averaged rest-mass density.  Such a correlation was suggested in Refs.~\cite{Takami15, Bauswein12a}.   This figure shows that in the later 3 kHz post-merger phase, the  $f_{peak}$ frequencies cluster somewhere between a soft and stiff EOS.  Hence, observing such a transition in the PSD between soft to stiffness, as evidenced in the different behaviors of $f_{max}$ and $f_{peak}$, may indicate the formation of quark matter.  Once the existence of quark matter is confirmed the amplitude of the PSD at $f_{peak}$ might be suggestive the strength of the quark couplings.

\section{Acknowledgements}

Work at the Center for Astrophysics of the University of Notre Dame is supported by the U.S. Department of Energy under Nuclear Theory Grant No. DE-FG02-95-ER40934. This research was supported in part by the Notre Dame Center for Research Computing through the high performance computing resources. H.I.K. graciously thanks Jinho Kim and Chunglee Kim for continuous support. The work of H.I.K. was supported by Basic Science Research Program through the National Research Foundation of Korea (NRF) funded by the Ministry of Education through the Center for Quantum Spacetime (CQUeST) of Sogang University (Grant No. NRF-2020R1A6A1A03047877). This research used resources of the Oak Ridge Leadership Computing Facility at the Oak Ridge National Laboratory, which is supported by the Office of Science of the U.S. Department of Energy under Contract No. DE-AC05-00OR22725.
\textsl{Notice}: This manuscript has been authored by UT-Battelle, LLC under Contract No. DE-AC05-00OR22725 with the U.S. Department of Energy.  The publisher, by accepting the article for publication, acknowledges that the U.S. Government retains a non-exclusive, paid up, irrevocable, world-wide license to publish or reproduce the published form of the manuscript, or allow others to do so, for U.S. Government purposes. The DOE will provide public access to these results in accordance with the DOE Public Access Plan (http://energy.gov/downloads/doe-public-access-plan). 

Software used was as follows \texttt{The Einstein Toolkit} (Ref. \cite{ET}; \url{https://einsteintoolkit.org}), \texttt{LORENE} (Refs. \cite{LORENE, LORENE2}), \texttt{PyCactus} (\url{https://bitbucket.org/GravityPR/pycactus}), and \texttt{TOVsolver} (\url{https://github.com/amotornenko/TOVsolver}).

%tidal deformability \\
%the upper limit estimated by the LIGo is Lambda1.4 < 800 at a 90\% confidence level. --> radius constraint R1.4 < 13.6km. LIGOnew R1.4=11.9+pm1.4km

%fmax vs Lambda \\ refs are given below
%71, 368-373
%fpeak vs ... \\ refs are given below
%fpeak vs Lambda 71, 373, 374 (Fig.3 of Bauswein2019(=109), unequal mass 105=369 ,106=371. cf 370=Takami2015)
%fpeak vs radius compactness, 120,122,369-371,373,375-383
%latest: lioutas21

%\bibliography{references}

\begin{thebibliography}{99}

\bibitem{Kedia22} A. Kedia, H. I. Kim, I.-S.  Suh, and G. J. Mathews, Phys. Rev. D 106, 103027 (2022).
\bibitem{Baiotti19} L. Baiotti, Prog. Part. Nucl. Phys. 109, 103714 (2019).
\bibitem{Radice20} D. Radice, S. Bernuzzi, and A. Perego, Annu. Rev. Nucl. Part. Sci. 70, 95 (2020).

\bibitem{LIGO-GW170817} B.P. Abbott \textit{et al}. (LIGO Scientific Collaboration and Virgo Collaboration), Phys. Rev. Lett. 119, 161101 (2017).
\bibitem{LIGO-GW170817eos} B.P. Abbott \textit{et al}. (LIGO Scientific Collaboration and Virgo Collaboration), Phys. Rev. Lett. 121, 161101 (2018).
\bibitem{Lattimer12} J. M. Lattimer, Annu. Rev. Nucl. Part. Sci., 62, 485 (2012).

\bibitem{Miller19} M. C. Miller \textit{et al}., Astrophys. J. Lett., 887 L24 (2019).
\bibitem{Riley21} T. E. Riley, \textit{et al}., Astrophys. J. Lett., 918 L27 (2021).
\bibitem{Miller21} M. C. Miller \textit{et al}., Astrophys. J. Lett., 918 L28 (2021).

\bibitem{Flanagan08} E. E. Flanagan and T. Hinderer, Phys. Rev. D 77, 021502 (2008).
\bibitem{Hinderer08} T. Hinderer, Astrophys. J. 677, 1216 (2008).
\bibitem{Read09tidal} J. S. Read, C. Markakis, M. Shibata, \textit{et al}., Phys. Rev. D 79, 124033 (2009).



%\bibitem{LIGO2} B. P. Abbott, \textit{et al}. "Tests of general relativity with GW170817." Phys. Rev. Lett. 123.1 (2019): 011102.
%\bibitem{Damour} T. Damour, M. Soffel, and C. Xu, Phys. Rev. D 45, 1017 (1992).


%Additional Refs \\
%refs 2-8 of Most19
\bibitem{Most18} E. R. Most, L. R. Weih, L. Rezzolla, and J. Schaffner-Bielich, Phys. Rev. Lett. 120, 261103 (2018).
\bibitem{Annala18} E. Annala, T. Gorda, A. Kurkela, and A. Vuorinen, Phys. Rev. Lett. 120, 172703 (2018).
\bibitem{Kurkela10} A. Kurkela, P. Romatschke, and A. Vuorinen, Phys. Rev. D 81, 105021 (2010).
\bibitem{Komoltsev22} O. Komoltsev and A. Kurkela, Phys. Rev. Lett. 128, 202701 (2022).
\bibitem{Fujimoto22} Y. Fujimoto, K. Fukushima, Y. Hidaka, A. Hiraguchi, and K Iida, Phys. Lett. B835, 137524 (2022).
\bibitem{Gorda22} T. Gorda, O. Komoltsev, and A. Kurkela, arXiv:2204.11877 (2022).
\bibitem{Tootle22} S. Tootle, C. Ecker, K. Topolski, T. Demircik, M. J\"rvinen, and L. Rezzolla, arXiv:2205.05691 (2022).
\bibitem{Weih20} L. R. Weih, M. Hanauske, and L. Rezzolla, Phys. Rev. Lett. 124, 171103 (2020).
\bibitem{Radice17} D. Radice, Astrophys. J. Lett. 838, L2 (2017).
\bibitem{Bauswein19} A. Bauswein, N.-U.F. Bastian, D. B. Blaschke, K. Chatziioannou, J. A. Clark, \textit{et al}., Phys. Rev. Lett., 122, 061102 (2019).
\bibitem{Blacker20} S. Blacker, N.-U.F. Bastian, A. Bauswein, \textit{et al}., Physical Review D 102, 123023 (2020).
\bibitem{Breschi19} M. Breschi, S. Bernuzzi, F. Zappa, M. Agathos, A. Perego, D. Radice, and A. Nagar, Phys. Rev. D 100, 104029 (2019).
\bibitem{Bauswein12a} A. Bauswein and H. T. Janka, Phys. Rev. Lett. 108, 011101 (2012).
\bibitem{Hotokezaka13}K. Hotokezaka, K. Kiuchi, K. Kyutoku, T. Muranushi, Y.-I. Sekiguchi, M. Shibata, and K. Taniguchi, Phys. Rev. D 88, 044026 (2013).
\bibitem{Bernuzzi14} S. Bernuzzi, A. Nagar, S. Balmelli, T. Dietrich, and M. Ujevic, Phys. Rev. Lett. 112, 201101 (2014).
\bibitem{Rezzolla16} L. Rezzolla and K. Takami, Phys. Rev. D 93, 124051 (2016).
\bibitem{Zappa18} F. Zappa, S. Bernuzzi, D. Radice, A. Perego, and T. Dietrich, Phys. Rev. Lett. 120, 111101 (2018).
\bibitem{Most20} E. R. Most, L. J. Papenfort, V. Dexheimer, M. Hanauske, H. Stoecker, and L. Rezzolla, Eur. Phys. J. A 56, 59 (2020).
\bibitem{Most19} E. R. Most, L. J. Papenfort, V. Dexheimer, M. Hanauske, S. Schramm, H. St\"ocker, and L. Rezzolla, Phys. Rev. Lett. 122, 061101 (2019).

\bibitem{Liebling21} S. L. Liebling, C. Palenzuela, and L. Lehner, Classical Quantum Gravity 38, 115007 (2021)
\bibitem{Prakash21} A. Prakash, D. Radice, D. Logoteta, A. Perego, V. Nedora, \textit{et al}., Phys. Rev. D 104, 083029 (2021).

\bibitem{Bauswein12b} A. Bauswein, H.-T. Janka, K. Hebeler, and A. Schwenk, Phys. Rev. D 86, 063001 (2012).
\bibitem{Bauswein20} A. Bauswein, S. Blacker, V. Vijayan, N. Stergioulas, K. Chatziioannou, \textit{et al}., Phys. Rev. Lett., 125, 141103 (2020).

% Issue of Phase transition

\bibitem{Pisarski16} R. Pisarski and S. V. Vladimir, Phys. Rev. D 94, 034015 (2016).
\bibitem{Steinheimer11} J. Steinheimer and S. Schramm, Phys. Lett. B 696, 257 (2011).
\bibitem{Hatsuda06} T. Hatsuda, M. Tachibana, N. Yamamoto, and G. Baym, Phys. Rev. Lett. 97, 122001 (2006).
\bibitem{Aoki06} Y. Aoki, G. Endr\'odi, Z. Fodor, S. D. Katz, and K. K. Szab\'o, Nature (London) 443, 675 (2006).
\bibitem{Baym18} G. Baym, T. Hatsuda, T. Kojo, \textit{et al}., Rep. Prog. Phys. 81, 056902 (2018)
\bibitem{Baym19} G. Baym, S. Furusawa, T. Hatsuda, T. Kojo, and H. Togashi, Astrophys. J. 885, 42 (2019)
% back to the PT issue
\bibitem{Kojo21} T. Kojo, G. Baym, T. Hatsuda, arXiv:2111.11919 [astro-ph.HE], (2021).
\bibitem{Kronfeld12} A. S. Kronfeld, Annu. Rev. Nucl. and Part. Sci. 62, 265 (2012).
\bibitem{Demorest10}P. B. Demorest, T. Pennucci, S. M. Ransom, M. S. E. Roberts, and J. W. T. Hessels, Nature (London) 467, 1081 (2010).
\bibitem{Antoniadis13} J. Antoniadis, P. C. C. Freire, N. Wex, T. M. Tauris, R. S. Lynch \textit{et al}., Science 340, 448 (2013).
\bibitem{Fonseca21} E. Fonseca, H.T. Cromartie, T.T. Pennucci, P.S. Ray, A.Y. Kirichenko \textit{et al}., Astrophys. J. Lett. 915, L12 (2021).
\bibitem{Sathyaprakash12} B. Sathyaprakash et al., Class. Quant. Grav. 29, 124013 (2012).
\bibitem{Abbot17} B. P. Abbott et al. (LIGO Scientific), Class. Quant. Grav. 34, 044001 (2017).
\bibitem{Hatsuda94} T. Hatsuda, and T. Kunihiro, Phys. Rep. 247, 221 (1994).
\bibitem{Alford08} M. G. Alford, A. Schmitt, K. Rajagopal, and T. Sch\"{a}fer, Rev. Mod. Phys., 80, 1455 (2008).
\bibitem{Togashi13} Togashi, H., and Takano, M. Nucl. Phys. A902, 53 (2013)
\bibitem{Togashi17} Togashi, H., Nakazato, K., Takehara, Y., et al. 2017, Nucl. Phys A961, 78 (2017).
\bibitem{Akmal98} A. Akmal, V. R. Pandharipande, D. G. Ravenhall, Phys. Rev., C 58, 1804 (1998).

\bibitem{SLy} E. Chabanat, P. Bonche, P. Haensel, J. Meyer, and R. Schaeffer, Nucl. Phys. A 635, 231 (1998).
\bibitem{GNH3} N. K. Glendenning, Astrophys. J. 293, 470 (1985).
\bibitem{Nambu61a} Y. Nambu and G. Jona-Lasinio, Phys. Rev. 122 (1961) 345.
\bibitem{Nambu61b} Y. Nambu and G. Jona-Lasinio, Phys. Rev. 124 (1961) 246.
\bibitem{Buballa05} Buballa, M. 2005, Phys. Rep., 407, 205
\bibitem{Read09} J. S. Read, B. Lackey, J. L. Friedman, and B. Owen, Phys. Rev. D 79, 124032 (2009).
\bibitem{ET} Z. Etienne {\it et al}, (2021), The Einstein Toolkit (The ``Lorentz" release, ET\_2021\_05). To find out more visit \url{doi.org/10.5281/zenodo.4884780}.
\bibitem{NOK} T. Nakamura, K. Oohara, and Y. Kojima, Prog. Theor. Phys. Suppl. 90, 1 (1987).
\bibitem{BSSN1} M. Shibata and T. Nakamura, PRD, 52, 5428 (1995).
\bibitem{BSSN2} T. W. Baumgarte and S. L. Shapiro, PRD, 59, 024002 (1999).
\bibitem{BSSN3} M. Alcubierre, B. Br\"ugmann, T. Dramlitsch, J. A. Font, P. Papadopoulos, E. Seidel, N. Stergioulas, and R. Takahashi, Phys. Rev. D 62, 044034 (2000).
\bibitem{BSSN4} M. Alcubierre, B. Br\"ugmann, P. Diener, M. Koppitz, D. Pollney, E. Seidel, and R. Takahashi, Phys. Rev. D 67, 084023 (2003).
\bibitem{Hawke05} I. Hawke, F. L\"offler, and A. Nerozzi, Phys. Rev. D 71, 104006 (2005).
\bibitem{Mosta14} P. M\"osta, B. C. Mundim, J. A. Faber, R. Haas, S. C. Noble, T. Bode, F. L\"ffler, C. D. Ott, C. Reisswig, and E. Schnetter, Classical Quantum Gravity 31, 015005 (2014).
\bibitem{Baiotti} L. Baiotti, I. Hawke, P.J. Montero, F. L\"offler, L. Rezzolla {\it et al}., Phys. Rev. D 71, 024035 (2005).
\bibitem{Valencia1} F. Banyuls, J. A. Font, J. M. Ibanez, J. M. Marti, and J. A. Miralles, Astrophysical Journal 476, 221 (1997).
\bibitem{Valencia2} J. A. Font, Numerical Hydrodynamics and Magnetohydrodynamics in General Relativity, Living Rev. Relativity 11 (2008).

%\bibitem{WENOZ} R. Borges, M. Carmona, B. Costa, and W. Don, J. Comput. Phys. 227, 3191 (2008).

\bibitem{LORENE} E. Gourgoulhon, P. Grandcl\'ement, K. Taniguchi, J.-A. Marck, and S. Bonazzola, Phys. Rev. D 63, 064029 (2001).
\bibitem{LORENE2} P. Grandclement and J. Novak (2007), Living Reviews in Relativity 12. 1 (2009).
\bibitem{Carpet1} E. Schnetter, S. H. Hawley, and I. Hawke, Classical Quantum Gravity 21, 1465 (2004).
\bibitem{Carpet2} E. Schnetter, P. Diener, E. N. Dorband, and M. Tiglio, Classical Quantum Gravity 23, S553 (2006).
%\bibitem{Carpet3} http://www.carpetcode.org/.
\bibitem{Pietri16} R. De Pietri, A. Feo, F. Maione, \& F. L\"offler, Physical Review D, 93(6), 064047 (2016).
\bibitem{Takami15} K. Takami, L. Rezzolla, L. Baiotti, Phys. Rev. D 91, 064001 (2015).
\bibitem{Read13} J.S. Read, L. Baiotti, J. D. E. Creighton, J. L. Friedman, B. Giacomazzo, et al., Phys. Rev. D 88, 044042 (2013).


\end{thebibliography}

%\end{document}

% end of file template.tex

%\iffalse

\end{document}